\documentclass[ preprintnumbers,twocolumn]{revtex4}
\usepackage{graphicx}
\usepackage{epsfig}
\usepackage{bm}
\usepackage{amsmath}
\usepackage{amssymb}
\usepackage{epstopdf}
\begin{document}

\title{Tracking energy fluctuations from fragment partitions in 
the Lattice Gas model}
\author{ F. Gulminelli}
\altaffiliation{member of the Institut Universitaire de France}
\affiliation{
LPC Caen (IN2P3 - CNRS / EnsiCaen et Universit\'{e}), F-14050
Caen C\'{e}dex, France} 
\author{Ph.Chomaz}
\affiliation{GANIL ( DSM - CEA / IN2P3 - CNRS),
B.P.5027, F-14076 Caen C\'edex 5, France} 
\author{M.D'Agostino} 
\affiliation{Dipartimento di Fisica and INFN, Bologna, Italy}
 
\begin{abstract}
Partial energy fluctuations are known tools 
to reconstruct microcanonical heat capacities. 
For experimental applications, approximations have been developed 
to infer fluctuations at freeze out from the observed fragment partitions. 
The accuracy of this procedure as well as the underlying independent fragment 
approximation is under debate already at the level of equilibrated systems.  
Using a well controlled computer experiment, the Lattice Gas model, 
we critically discuss the thermodynamic conditions under which
fragment partitions can be used to reconstruct the thermodynamics
of an equilibrated system. 
\end{abstract}

\maketitle

\section{Introduction}

The observation of abnormal fluctuations in nuclear
multifragmentation and its possible connection to a negative
heat capacity\cite{michela} has raised much interest and discussions 
in the last years\cite{claudio_cc,moretto,thirring,campi,farizon,zero}.

For a microcanonical ensemble, it has been proposed in ref. \cite{NegC-theo} 
that the kinetic energy fluctuation $\sigma_{K}^{2}/T^{2}$ can be used 
to reconstruct the heat capacity even in the context of phase transitions
and for small systems.
The accuracy of the fluctuation expression has been 
successfully tested on numerical experiments on the 
liquid gas phase transition, using the microcanonical 
Lattice Gas model\cite{prl00} and 
molecular dynamics simulations with a Lennard-Jones potential
\cite{campi,cherno}.

From the experimental point of view, it has been proposed to use  
the clusters asymptotically detected in nuclear multifragmentation 
reactions to backtrace  the fluctuations of the total 
energy partitioning at freeze-out 
\cite{michela}. 
The robustness of the experimental procedure 
has been tested in ref. \cite{palluto}. 
In particular, statistical models have been used 
to generate events, then analyzed using the experimental procedure 
and a good reproduction of the model heat capacities has been 
reported \cite{palluto}.  Using molecular dynamics at equilibrium 
and recognizing fragments through the Hill algorithm\cite{hill}  
the authors of ref. \cite{campi}
have come to opposite conclusions criticizing the independent 
fragment hypothesis on which is based the experimental method 
as well as the zero temperature approximation for the fragment binding 
used to evaluate the fragment internal interaction energy.  
The model of ref.\cite{campi} has also been studied 
in ref.\cite{cherno},
where the dynamics of the expansion was explicitly included.
The result was that the fluctuations of dense systems 
were strongly modified by the dynamics and that 
only the latest phase of the expansion, 
after a freeze-out corresponding to dilute configurations, 
can be accessed from 
the observation of asymptotic partitions.

More generally, the question of the thermodynamic conditions
under which the energy partitioning of a 
small fragmenting system can be retraced from the measured fragment
sizes and kinetic energies, raises important  
questions for the whole field of nuclear thermodynamics. 
Indeed the independent fragment hypothesis is not only needed 
to reconstruct partial energy fluctuations\cite{michela} but is 
also necessary for any other quantitative estimation of the 
nuclear phase diagram\cite{elliott}.
Moreover only if the fragmenting source 
can be approximated by an ensemble of non- (or weakly-) interacting
nuclear clusters, the statistical models\cite{statistical} 
that have successfully reproduced
heavy ion data since two decades can be theoretically justified.

To contribute to this debate, 
we want to address the problem of the independent fragment
hypothesis  and of the fragment energetics 
at equilibrium in the framework 
of a well controlled exact numerical model, the Lattice Gas
model\cite{prl00}.
We will show in this article that, in the Lattice Gas model, 
the system heat capacity can be well estimated from fragment sizes
at all energies and for all pressures $p/p_c \lesssim 1/3$
almost independent of the parametrization adopted 
for the fragment energies.
For higher pressures this approximation tends to break down, 
but the estimation of fluctuation stays at the 30\% 
accuracy 
level even in the supercritical regime. 

\section{Lattice gas results}

It has been proposed in \cite{NegC-theo} to use the partitioning of a fixed
total energy ($E_{tot}$) into kinetic ($K$) and configurational ($V$)
energies, $E_{tot}=K+V$, 
in order to look for entropy curvature anomalies. 
Indeed, for classical systems with momentum independent interactions, 
because of the microstates equiprobability the microcanonical 
distribution of $K$ at a fixed energy $E_{tot}$ reads 
\[
P_{E}\left( K\right) =\exp \left( S_{K}\left( K\right) +S_{V}\left( V\right)
-S_{tot}\left( E_{tot}\right) \right) 
\]
where  $S_{K}, S_{V},S_{tot}$ are the kinetic, potential, and 
total entropies. If 
this 
distribution is
normal, a gaussian approximation can be performed leading to analytic
expressions relating temperatures and heat capacities to the observed
moments of the distribution. When only the leading order is kept, we
recover the simplest expression used in ref. \cite{signa-first} 
to relate the total microcanonical heat capacity $C$ to the kinetic one $C_K$ and the kinetic energy fluctuation $\sigma_K$ scaled by the system temperature $T$ 
\cite{footnote} 
\begin{equation}
C=C_{K}\left( 1-\frac{\sigma _{K}^{2}}{C_{K}T^{2}}\right) ^{-1}  \ \ \ .
\label{EQ:1} 
\end{equation}
Even if in practice it seems that Eq. (\ref{EQ:1}) is often accurate enough,
the validity and the accuracy of this approximate expression   
should always be checked 
by controlling the actual distribution and when the distribution is normal by
evaluating the corrections. In our work  this has been
done both in theory\cite{NegC-theo} and experiments\cite{palluto}. 
The accuracy of the
simple expression (\ref{EQ:1}) to infer the heat capacity 
even in the vicinity of a phase transition, 
or worse of a critical point
is also verified in refs.\cite{campi,cherno}. 
 
The application to experimental data of the idea proposed in \cite{NegC-theo}
requires the development of different tools. 
Let us first briefly recall the problem of the potential energy 
determination for an ensemble of fragments.
If we look at a system of $A$ interacting particles 
as a system of $M_f$ clusters
(including monomers), 
the potential energy $V=\sum_{i<j}^A v_{ij}$ can be written as
$V=\sum_{f=1}^{M_f}V_f + \sum_{f<g}^{M_f} V_{fg}$ where  
$V_f=\sum_{i<j\\ \epsilon f}v_{ij}$ and $V_{fg}=
\sum_{i\epsilon f,\\ j\epsilon g,\\ f\neq g}v_{ij}$ 
are the intrafragment and 
interfragment components respectively. 
In the experimental analysis of ref.\cite{michela} 
the only interaction considered among the different 
fragments is the Coulomb force
because of its long range nature. The interfragment nuclear 
force is thus neglected following the argument that an important nuclear 
interaction is incompatible with the freeze-out concept.  
Concerning the evaluation of the 
intrafragment potential energy $V_f=\sum_{i<j\\ \epsilon f}v_{ij}$, 
this latter is approximated in the experimental analysis\cite{michela}
by the tabulated ground state nuclear energies.
Both these approximations are used in many 
other thermodynamic studies of multifragmentation
\cite{eos,elliott} and in all macroscopic 
statistical models\cite{statistical}.

\subsection{Independent Liquid drop approximation}\label{first}

In order to check the quality of these approximations,
we shall  use an exact numerical 
experiment. 
This restricts studies to classical systems.
At variance with nuclear systems which are 
liquid in their ground states,  the ground states of classical models
are solid and so present an extra binding.
To avoid this difficulty, we 
may
approximate 
the interaction energy $V$   
as a sum of independent liquid drops contributions
\begin{equation}
Q_{LD}=\sum_{f=1}^{M_f}B_f=\sum_{f=1}^{M_f}a_V A_f+a_S A_f^{2/3}
\label{ldm}
\end{equation}
where $A_f$ is the size of cluster $f$ and $M_f$ is the total number of
clusters.
The parameters $a_V$, $a_S$ 
can then be
fitted to reproduce the cluster binding 
at low but finite temperature to avoid the peculiarities of the ground states 
of such classical systems. 

In this article we present a study based on exact numerical experiments of 
the 3D Lattice gas model\cite{prl00}. 
The ground states of such a model are cubes and, as discussed above, to estimate 
the internal interaction energy of the liquid fragments  
avoiding the extra binding of those peculiar cubic configurations 
we shall use low temperature simulations.    
Using canonical simulations with temperatures around 1/3 of the  critical 
temperature (or microcanonical ones with energies around $-1.5 \epsilon$)
leads to $a_V=-2.86\epsilon$, $a_S=2.73\epsilon$, 
where $\epsilon$ 
is the lattice coupling (see discussion of Fig 1).   

For comparison the choice
$a_V=-3\epsilon$, $a_S=3\epsilon$ corresponds to 
large
cubic
clusters, while $a_V=-3.06\epsilon$, $a_S=3.35\epsilon$ leads 
to a good description of the zero temperature 
clusters
in the size range 
$2<A_f<60$.

In the following calculations the coefficients $a_V=-2.86\epsilon$, 
$a_S=2.73\epsilon$ will be kept constant, 
and we will come back to the influence of the parameters values in the 
last section.

\begin{figure}[htbp]
\begin{center}
\includegraphics[width=9cm]{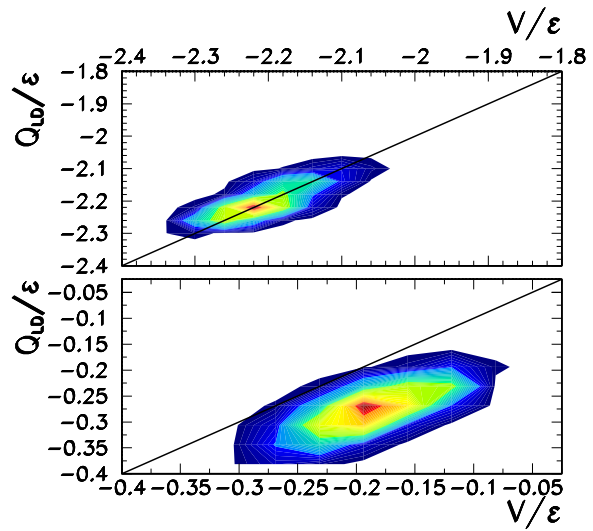}
\end{center}
\vskip -1.cm
\caption{\it Correlation between the exact interaction energy $V$
and its approximation from fragment sizes $Q_{LD}$ (see text)  
 at an average density $\rho/\rho_0=0.0135$ and a temperature 
$T/T_c=0.29$ (upper part), and $T/T_c=0.57$ (lower part).
The liquid drop parameters for the fragment binding 
energy are fixed as 
$a_V=-2.86\epsilon$, $a_S=2.73\epsilon$.}
\label{fig:1}
\end{figure}

\begin{figure}[htbp]
\vskip -2.cm
\begin{center}
\includegraphics[width=9cm]{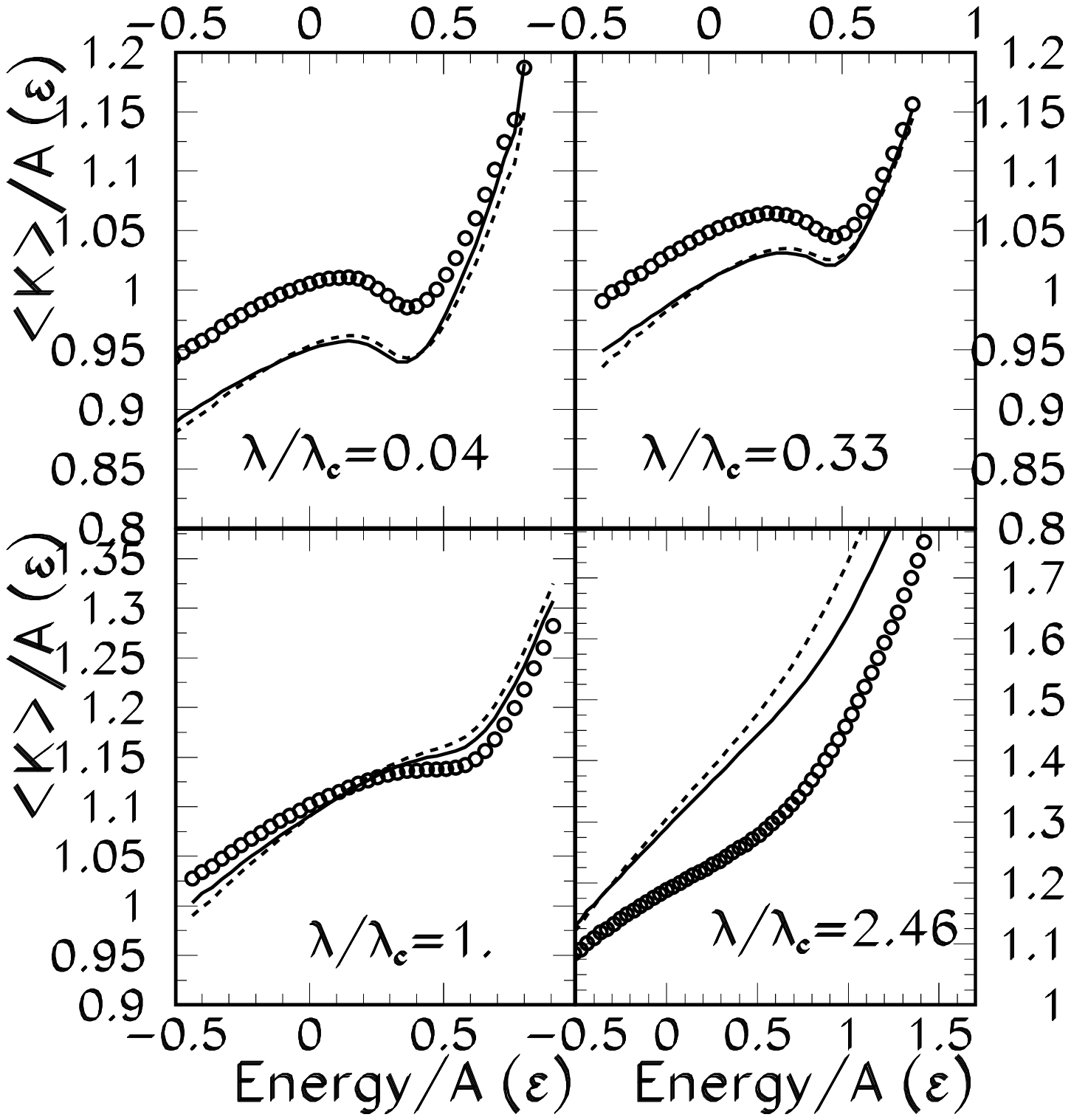}
\end{center}
\vskip -1.cm
\caption{\it Reconstruction of the average kinetic energy
(solid lines) as a function of total energy 
from fragment sizes in the Lattice Gas model at different
pressures. Dashed lines: liquid drop parameters from table 1. 
Symbols: liquid drop parameters
fixed from the low temperature low density phase as in
figure 1.}
 \label{fig:2}
\end{figure}
 
\begin{figure}[htbp]
\vskip -2.cm
\begin{center}
\includegraphics[width=9cm]{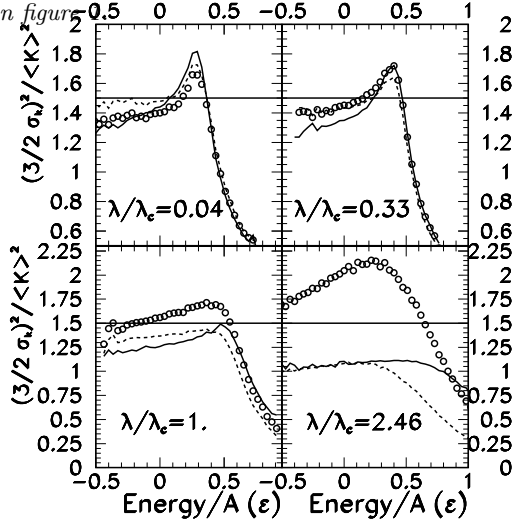}
\end{center}
\vskip -1.cm
\caption{\it Reconstruction of the normalized kinetic energy
fluctuation (solid lines) as a function of total energy 
from fragment sizes in the Lattice Gas model at different
pressures. Dashed lines: liquid drop parameters from table
1. Symbols: liquid drop parameters
fixed from the low temperature low density phase as in
figure 1.}
\label{fig:3}
\end{figure}

We have 
first tested the accuracy of the approximation (\ref{ldm})
for the total interaction energy $V$ 
in a low density case (108 particles in a cubic box of volume
$L^3=8000$) where the interfragment energies can be safely neglected.
The upper part of figure \ref{fig:1} shows the correlation between 
the exact interaction energy
and its independent fragment approximation eq.(\ref{ldm})
in a canonical simulation at a temperature $T/T_c=0.29$.
The corresponding average total energy is $-1.74\epsilon$,
well below the backbending region.
The good linearity of the plot 
(the $Q_{LD}/V$-correlation coefficient 
is $c=0.77$
) 
shows that indeed in such a low density configuration
the interaction energy can be calculated within a liquid drop approximation. 

Even in the independent fragment picture, the liquid drop
coefficients are expected to change with temperature reflecting the 
internal excitation of the produced clusters.
This can be seen in the lower part of figure \ref{fig:1}, 
which shows a calculation
for the same very diluted system at a temperature well above the 
transition temperature. It is clear that the quality of the correlation
does not 
significantly change
with the temperature 
(here the $Q_{LD}/V$-correlation coefficient is $c=0.78$), 
but the use of the low temperature mass formula leads to a systematic 
$\approx 20 \%$ overestimation of the fragment binding 
that can affect the thermodynamic analyses.

\subsection{The choice of the statistical ensemble}

This point should be further explored 
looking at direct effects of the considered approximation on 
thermodynamical quantities, i.e. on  ensemble averaged 
observables such as averages and variances.
Since the different statistical ensembles are not equivalent in finite
systems, the statistical ensemble has to be specified.
In the following we have chosen to perform calculations 
in the microcanonical 
"isobar" ensemble characterized by 
the two state variables $(E,\lambda)$,
the total energy and the Lagrange multiplier imposing the average 
volume $V$, respectively\cite{prl00}.
Statistical averages are calculated as

\begin{equation}
\langle A \rangle_{E,\lambda} = \frac{\sum_{(n)}A^{(n)}
exp\left ( -\beta E^{(n)} -\lambda V^{(n)}\right )
\delta\left ( E^{(n)}-E\right)}
{\sum_{(n)}exp\left ( -\beta E^{(n)} -\lambda V^{(n)}\right )
\delta\left ( E^{(n)}-E\right)}
\end{equation}

where $A$ is a generic observable ($A=K$ allows to 
compute the microcanonical 
temperature, $A=K^2$ provides the heat capacity), 
the sum runs over the system
microstates, and the average volume is defined 
through the one body 
observable 

\begin{equation}
V=\frac{4\pi}{3L^3}\sum_{i=1}^{L^3} r_i^3 n_i
\end{equation}

where $n_i=0,1$ is the occupation of the $i-th$ lattice site.

Different reasons motivate the choice of this ensemble.
First, this is the ensemble in which the liquid gas phase 
transition is associated to a negative heat capacity
up to the critical point\cite{houches}.
Moreover in the actual analysis of heavy ion experiments data are sorted 
in excitation energy bins, i.e. approximate realizations of 
microcanonical ensembles. On the other side the system
extension is only imposed by the freeze out requirement without any 
boundary condition. We have recently shown that such a physical situation 
is accounted in a thermostatistical coherent way only if the system size
is imposed through a lagrange parameter $\lambda$, with $P=\lambda/\beta$ 
having the dimension (and the physical role) of a constraining
pressure\cite{annals}.
Finally it is interesting to notice that if $\lambda=0$ this "isobar"
ensemble is equivalent to the isochore ensemble used in section \ref{first}
if the volume is large enough such that the boundary conditions become
irrelevant. 
  
\subsection{Averages and fluctuations}
Figure \ref{fig:2} 
shows the estimated 
average kinetic energy $E-\sum_{f}B_{f}$ as a function of the total
energy with different constraints  on the system volume\cite{prl00}, 
corresponding to different pressures.
The shape of the exact caloric curve (solid line) 
is nicely reproduced by the 
independent "cold" liquid drop approximation (open symbols) 
for all volume constraints.
Using the temperature backbending to define the phase transition, 
we see   
that the coexistence zone as well as the critical point
can be well estimated from the unique knowledge of the fragment 
partitions. However, the actual value of the temperature 
shows a systematic shift. The fact that this shift is also present 
in the pure high temperature low density gas phase 
(figure \ref{fig:1}) suggests that it 
may be due  
to a temperature dependence of the fragment internal
energy which is not accounted 
for
in our liquid drop parametrization.

\begin{figure}[htbp]
\vskip -2.cm
\begin{center}
\includegraphics[width=9cm]{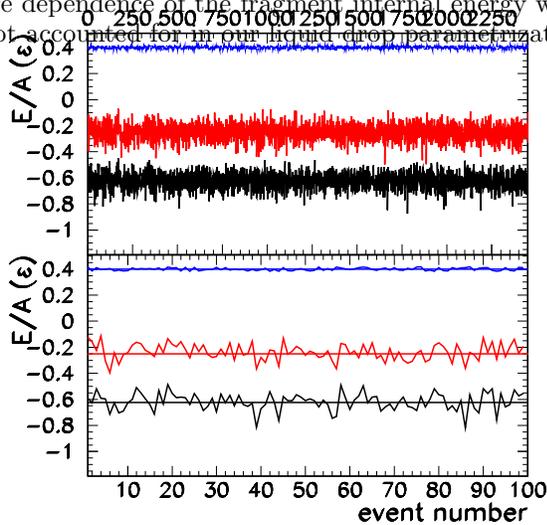}
\end{center}
\vskip -1.cm
\caption{ \it
Upper panel: partial energies of an ensemble of events 
at a pressure $\lambda/\lambda_c=.33$ and energy 
$.38\epsilon < E_{tot}/N < .42\epsilon$
in the middle of the coexistence region.
From bottom to top:
-interaction energy per particle;
-its independent fragment approximation (LD parameters as in figure 1)
  (shifted by 0.4);
-total energy per particle.
The horizontal lines give the average values of the corresponding energies.
Lower panel: same as the upper panel on an extended scale.}
 \label{fig:0}
\end{figure}

The effect on partial energy fluctuations is shown in figure \ref{fig:3} 
for the same thermodynamic conditions as in figure \ref{fig:2}. 
We can see that the fluctuations tend to be overestimated, 
but since the bias on
the average value goes in the same direction, the normalized 
fluctuations are still reasonably reproduced almost
up to the critical point.

\begin{figure}[htbp]
\vskip -1.cm
\begin{center}
\includegraphics[width=9cm]{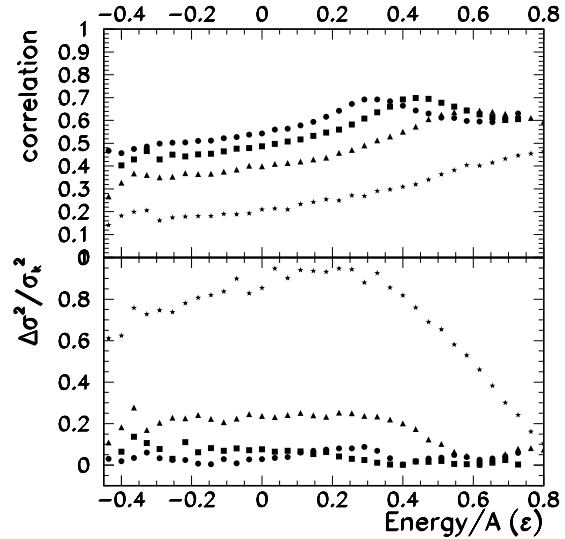}
\end{center}
\vskip -1.cm
\caption{\it  $Q_{LD}/V$-correlation coefficient (upper part)
and percentage error $| \sigma^2_Q-\sigma^2_K 
|/\sigma^2_K$ on the partial energy fluctuation from figure \ref{fig:3}
(lower part).
The liquid drop parameters are fixed as 
$a_V=-2.86\epsilon$, $a_S=2.73\epsilon$. Black points:$\lambda/\lambda_c=.04$,
squares:$\lambda/\lambda_c=.33$, triangles:$\lambda/\lambda_c=1$, stars:
$\lambda/\lambda_c=2.46$.  }
\label{fig:3bis}
\end{figure}

Even if the quantitative study of a statistical ensemble should be based 
on ensemble averages such as the one presented in the previous figures, 
it is instructive to look at 
a set of events corresponding to one of the above cases 
as
shown in Fig. \ref{fig:0}.  
This figure illustrates the fact that the actual 
interaction energy and its independent liquid-drop approximation 
have similar properties, averages and fluctuations. Moreover, the extended 
scale part shows that the two quantities are 
correlated. 


\subsection{Errors and correlations}

In order to quantify the quality of the liquid drop approximation 
let us look at 
ensemble averaged quantities which are indeed the information 
used to infer thermodynamical properties.    
One of the most important observables is the kinetic energy fluctuation. 
The corresponding percentage error $| \sigma^2_Q-\sigma^2_K 
|/\sigma^2_K$ 
is shown in the 
lower part of figure \ref{fig:3bis}. One observes that the error 
increases with the pressure but remains low for all energies  
up to the critical pressure. This error is below 
$10 \%$ up to 1/3 of the critical pressure and below $20 \%$ 
at the critical pressure. 
Moreover, when used to compute the heat capacity 
this error is partly compensated by the similar error on 
the kinetic energy 
used to deduce the temperature. This explains the good accuracy 
observed for the liquid drop approximation up to rather high 
temperatures and pressures. 

This good accuracy of the ensemble averaged quantities 
does not imply  nor require a similar accuracy on an 
event by event basis. In fact one expects stronger deviations when 
looking at a single event. One way to study this event by event 
accuracy of the liquid drop approximation is to study the correlation 
coefficient $c$. However, this is only a part of the discussion since 
the correlation coefficient is not sensitive 
to the magnitude of the fluctuation and only  
characterizes the link between the independent liquid drop
approximation and the actual internal energy.
In the considered case, this implies  that by 
construction the correlation coefficient
$c$ is independent of the value of the liquid drop parameters.   
The global trend of the correlation coefficient with pressure and energy
is presented in the upper part of figure \ref{fig:3bis}.
We can see that the best correlation is systematically observed 
around the fluctuation peak, where the size distribution is 
the broadest.  
At subcritical pressures, $c$ always exceeds
0.4, while the correlation decreases to around 0.2 in the supercritical
regime. 

It should be stressed that the idea to use,
in equation (1), the first moments (averages and fluctuations) 
of the event distribution was indeed to overcome the difficulty to 
get accurate information on an event by event basis. 
Though the best correlation obviously leads to the smallest error,
we can see that the relation between $c$ and $\Delta\sigma$ is not
trivial.
In particular a value of $c$ exceeding 0.4 comes out to be enough 
to keep the error below the 15\% level, independent of the energy.

It is also interesting to remark that the correlation coefficients
shown in figure \ref{fig:3bis} are systematically lower than the 
ones shown in the calculations of figure \ref{fig:1}. 
This is another illustration of the non equivalence of statistical
ensembles: in the canonical case the distributions are wider than in
the microcanonical one allowing a better correlation.
 
\subsection{Beyond the fixed cluster energy approximation}

The failure at increasing pressure shown by fig.\ref{fig:3} is interesting.
One may wonder whether the lack of reproduction is due to the breaking
down of the independent fragment approximation\cite{campi} in dense 
media, or whether the configurational contribution to the 
internal fragment excitation energy 
has to be taken into account
by a proper redefinition of the mass formula.
To answer to this question we have allowed a free variation of the 
liquid drop parameters according to table 1. The result, shown
by the dashed lines in figs.\ref{fig:2},\ref{fig:3}, 
is that both the caloric curve
and the fluctuations can be very precisely reproduced 
by the independent fragment approximation
in a wide range of temperatures and pressures
if the volume and surface coefficients are allowed to decrease 
with increasing excitation. 
Only at very high pressures and temperature, above the critical
point, the approximation appears to clearly break down: in this dense
configuration the objects identified 
as "fragments" by the cluster recognition algorithm 
have certainly little to share with physical isolated liquid drops. 
Indeed in this density regime
the Q value eq.(\ref{ldm}) is very poorly correlated with the 
interaction energy (see figure \ref{fig:3bis}). 

\begin{table}[h!]
{\small
\begin{center}
\begin{tabular}{|c|c|c|c|}
\hline $\frac{\lambda}{\lambda_c}$ & $\frac{<V>}{<V>_c}$& 
$\frac{a_v}{\epsilon}$& $\frac{a_s}{\epsilon}$\\
\hline
0.04& 2.00 & -2.76 & 2.66 \\
\hline 
0.33& 1.69 & -2.74 & 2.61 \\
\hline
1.00& 1.00 & -2.72 & 2.53 \\
\hline 
2.46& 0.53 & -2.64 & 2.18 \\
\hline 
\end{tabular}
\protect\caption{\label{table} \it Volume $a_V$ and surface $a_S$
effective liquid drop parameters (see text) 
allowing to reproduce the average
canonical configurational energy of 216 Lattice Gas particles
at the temperature corresponding to the maximal 
energy fluctuations, for different pressures normalized to the critical
pressure. The average volume occupied by the system 
divided by the critical volume is also given.}
\end{center}
}
\end{table}

\section{Discussion and possible experimental implications}

Let us now turn to the possible implications of these findings to 
heavy ion collisions experiments, and particularly to the determination
of the heat capacity from the fluctuations of asymptotic detected 
partitions.

In the application of eq.\ref{EQ:1} to the analysis of multifragmentation 
data several difficulties arise that have to be considered. 

First of all in the experimental case the only information available on 
the system is given by its cluster properties (sizes and kinetic energies).
This means that the total energy of the system 
is not evaluated from its microscopic constituents, 
but is also estimated from the measured kinetic energies 
and fragment sizes as $E_{tot}=Q_{LD}+K$. 

Moreover both $Q_{LD}$ and $K$ are time dependent variables, and in 
the data analysis the quantities at freeze out are extrapolated from
the asymptotic ones solely correcting for secondary evaporation and Coulomb
repulsion. In particular the configurational 
energy functional associated to each fragment at freeze out  
is 
assumed to be 
the one associated to $T=0$ and is not
pressure dependent. As we have seen in the previous section, 
this assumption is not valid for the Lattice Gas model: 
the presence of a 
cubic
 ground
state in this model leads to a mass formula that is not adequate
to reproduce the average cluster energy in the liquid phase.
The difference in energy is small compared to the liquid-vapour
latent heat, however we have seen that 
it can have important effects in the calculation
of the observables. This difficulty in principle should 
not arise in nuclear physics
where clusters are already liquid at T=0, however one may ask how much 
a possible modification with temperature and pressure of the fragments
energetics would influence the experimental analysis.  

Finally the assumption is explicitly made that at the freeze out time 
fragment partitions essentially reflect thermal equilibrium.

\begin{figure}[htbp]
\vskip -1.cm
\begin{center}
\includegraphics[width=9cm]{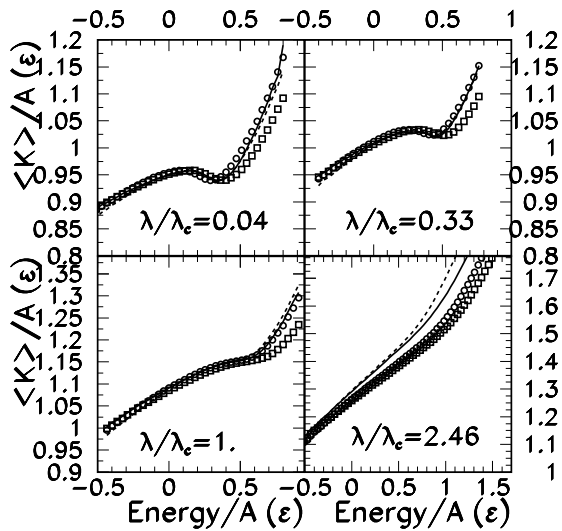}
\end{center}
\vskip -1.cm
\caption{\it Same as figure \ref{fig:2} above, 
but including the effect of the calorimetric estimation 
of total energy (see text). Squares: zero temperature liquid drop
parameters $a_v=-3\epsilon$,$a_s=3\epsilon$.} 
 \label{fig:4}
\end{figure}

\begin{figure}[htbp]
\vskip -1.cm
\begin{center}
\includegraphics[width=9cm]{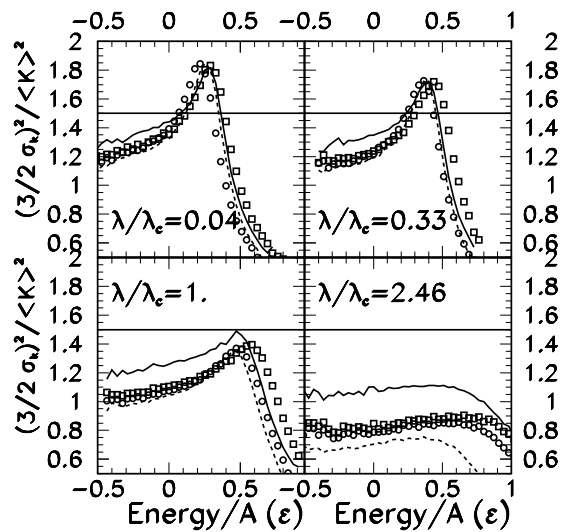}
\end{center}
\vskip -1.cm
\caption{\it Same as figure \ref{fig:3} above, 
but including the effect of the calorimetric estimation 
of total energy (see text). Squares: zero temperature liquid drop
parameters $a_v=-3\epsilon$,$a_s=3\epsilon$.}
\label{fig:5}
\end{figure}
 
\subsection{The thermodynamics of a gas of clusters}

Let us leave the difficult problem of equilibrium and time dependence
to the next subsection and start addressing the question of  
the quality of reconstruction of the thermodynamics 
of freeze-out, using accessible information (fragment sizes and total 
kinetic energies). In this matter, 
equilibrium studies provide a valuable testing ground, if the studied
thermodynamic conditions correspond to the freeze-out configurations 
of the model.

To quantify the uncertainty of the heat capacity reconstruction 
from the fragment information only, we have introduced reconstructed 
microcanonical statistical ensembles defined by the total energy 
constraint $E_{cal}=K+Q_{LD}$
at different pressures and 
with different prescriptions for the liquid drop parameters. 
Figures \ref{fig:4} and \ref{fig:5} show the resulting first and second 
moment of the $K$ distribution in bins of $E_{cal}$.
We can see that the systematic bias observed in fig.\ref{fig:2} disappears
if the same prescription for the interaction energy is used both
for the fluctuation and for the total energy of the system.
Up to the critical point, the precise parametrization of the liquid drop
energy does not change the results dramatically. 
In particular the zero temperature cubic solid cluster prescription 
$a_v=-3\epsilon$,$a_s=3\epsilon$ induces a spurious shift towards 
higher energies, but does not induce extra fluctuations.

These results may be qualitatively understood as follows. The  
error in the estimation of $V$ through $Q_{LD}$ in each 
event of a set at a given total physical energy 
induces an overall bias in the evaluation of $<E_{tot}>$, 
but also and more important, a 
spread in the total energy estimation. Both these effects are 
especially important at high pressures, where the correlation coefficient
between $Q_{LD}$ and $V$ is low (see fig.\ref{fig:3bis}).
The systematic bias in the estimation of the average total energy
depends directly on the value chosen for the liquid drop parameters, 
and leads to the shift towards higher energy observed in 
figs.\ref{fig:4},\ref{fig:5}.  
The width of the $E_{cal}$ distribution for each value of $E_{tot}$ 
is at the origin of the overestimation of the partial energy 
fluctuations observed 
in fig.\ref{fig:3}. When events are analyzed in bins of total 
estimated energy, these spurious energy fluctuations do not contribute
any more to the width of the $K$ distribution.
The remaining effects are due to the mixing in a given $E_{cal}$ bin
of events coming from different physical $E_{tot}$ values.
Since this mixing is approximately symmetric in energy, 
the average $<E_k>$ appears
to be not much affected and the main effect of mixing is to flatten out
the fluctuation curve. This is very similar to the results already 
reported for the analysis of events produced in macroscopic 
statistical models (SMM\cite{palluto} and SIMON\cite{vient}).

From figs.\ref{fig:4} and \ref{fig:5}
we can also see that, when calorimetry is taken into account, 
it is not possible any more to tune the liquid drop parameters
at high pressure such as to reproduce at the same time 
the correct average energy and fluctuation. 
This confirms again that the independent fragment hypothesis breaks down in 
the supercritical regime. 

\begin{figure}[htbp]
\vskip -1.cm
\begin{center}
\includegraphics[width=9cm]{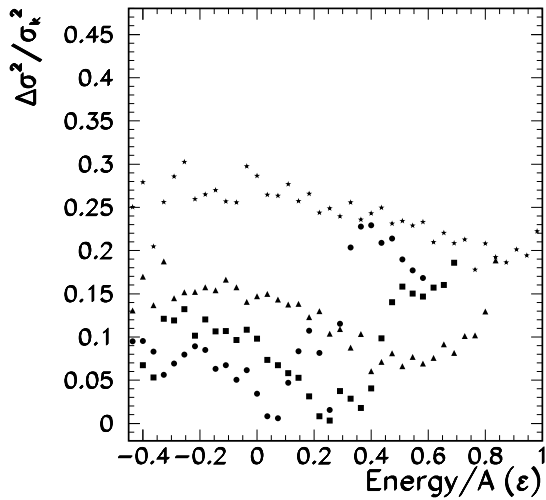}
\end{center}
\vskip -1.cm
\caption{\it Percentage error $| \sigma^2_Q-\sigma^2_K 
|/\sigma^2_K$ on the partial energy fluctuation from figure \ref{fig:5}.
The liquid drop parameters are fixed as 
$a_V=-2.86\epsilon$, $a_S=2.73\epsilon$. Black points:$\lambda/\lambda_c=.04$,
squares:$\lambda/\lambda_c=.33$, triangles:$\lambda/\lambda_c=1$, stars:
$\lambda/\lambda_c=2.46$.}
\label{fig:6}
\end{figure}

The error on the fluctuation estimation is reported in fig.\ref{fig:6}
for all the considered pressures and energies, using the fixed low
temperature liquid drop parameters $a_V=-2.86\epsilon$, $a_S=2.73\epsilon$.
The fact of taking into account the calorimetric shift does not 
increase this systematic error above the 20\% level at subcritical pressure.

The results of figs.\ref{fig:4},\ref{fig:5} 
suggest that if the freeze-out density 
is low enough, then the thermodynamic 
properties of the system can be deduced from the fragment partitions
using the independent fragment model with a fragment energy approximated by
its low temperature binding.
In the Lattice-gas model the validity condition of this simple approximation 
happens to be not so restrictive since it corresponds to an average volume 
$V\gtrsim 1.5 V_c$. 
If the freeze out density is higher, in the lattice-gas model the independent
fragment approximation tends to break down, and both the temperature
and the fluctuation are systematically underestimated.


\subsection{The influence of dynamical and quantum effects}

When dealing with reactions and trying to extract thermodynamic
information, the first question is the 
relevance of the freeze-out and equilibrium concepts at specific 
stages of the collision. 
From the theoretical point of view, this complex question, 
which is strongly debated since more then a decade,
can only be addressed by complete simulations of the 
reaction and 
critical analysis of the resulting time dependence.   
Such studies require to go beyond the equilibrium 
models used in the present article as well as in ref. \cite{campi}.
%
%
In a recent dynamical calculation with the 
Lennard-Jones hamiltonian, A.Chernomoretz et al.\cite{cherno} have 
addressed this question by looking at the time evolution of different
observables for a system initially thermalized in a confined box 
and subsequently freely evolving in the vacuum. Almost independent
of the initial energy and density, the average configurational 
energy and the associated fluctuations turn out to freeze 
when the density of the system is 
of the order of $\rho \approx 0.03$ in Lennard-Jones units, 
a region of the phase diagram well below
the critical point\cite{claudio_cc}. 
This study, thus, implies that 
only such diluted stages of the reaction
can be reconstructed from the asymptotic ("experimental") information. 
However, it is important
to stress that the density value at freeze out is a 
model dependent quantity and in particular it is 
correlated with the range of the force.
In the study of ref. \cite{cherno}, the configurations at  
freeze-out are
also characterized by an  interfragment energy sufficiently small 
$\sum_{f<g}^{M_f} V_{fg}(t)\approx 0$ for fragment partitions 
to be essentially frozen.  Conversely to the absolute 
value of the density, this condition might be a more robust definition 
of freeze out. It is also a first argument in favor of  
an independent fragment approximation.

Another consequence of the time dependence of a collisional process
is that the freeze out configuration may not be sufficiently close 
to a thermodynamic equilibrium to be described with statistical 
tools\cite{claudio_cc,indra}.  
The statistical nature of freeze-out
is indeed the key point in order to interpret the
reconstructed fluctuation in terms of heat capacity.
To estimate the distortions due to the out of equilibrium 
component one needs to know how much the 
averages and fluctuations deviate from the equilibrium 
values. 
Different verifications have been performed to estabilish 
this point\cite{palluto} in the experimental analyses of collisional 
data. In particular, the observation of the same behavior 
with different entrance channels is an interesting argument in favor of 
being close to an equilibrium. 
From a theoretical point of view,
this fundamental open question can only be addressed
through dynamical approaches and cannot be answered from 
statistical calculations as the one presented here.  

Finally it is important to stress that the results of classical 
models like the Lattice Gas (or the Lennard Jones analyzed in refs.
\cite{campi},\cite{cherno}) cannot be quantitatively applied 
to nuclear data because of the complete lack of quantum effects.
The ground state properties are not the only point where quantum
effects are expected to play an important role.
The actual value of the density at freeze out 
of a nuclear system can be 
very different from the one estimated from classical 
calculations\cite{cherno}, and the same is true for the limiting density 
that can be described through the independent fragment approximation
derived in this work. The consistent inclusion of quantum effects
in the statistical analyses of nuclear data
is an ambitious program that is only 
in its infancy\cite{houches} and  constitutes one of the greatest
theoretical challenges of heavy ion collisions in the next decades. 
 
 \ 
 
 \section{Conclusion}

To summarize, 
in this paper we have reconstructed configurational energy fluctuations
from the fragment partitions of a Lattice Gas model. 
As far as the first and second moments of collective observables 
are concerned, dilute systems in thermodynamic equilibrium 
can be accurately approximated by an ensemble of independent 
fragments. A unique liquid drop parametrization for the fragment 
binding energies is able to describe the thermodynamics of the 
system independent of the deposited energy or temperature.
A modification of the order of 10\% of the liquid drop parameters
does not modify the temperature and fluctuation in a sizeable way,
if the same parametrization is consistently employed for the 
determination both of the fluctuation and of the total energy.
For pressures of the order or above the critical point,
the independent fragment approximation tends to break down 
leading to a systematic underestimation of temperatures
and fluctuations at the 30\% level.

\acknowledgments{Discussions with F.Cannata are gratefully acknowledged.}


\begin{thebibliography}{99}
\bibitem{footnote} It is worthwhile noticing that, if
  eq.(\protect\ref{EQ:1}) has 
been derived in ref.\protect\cite{signa-first} only for the isochore
ensemble, the more general derivation of ref.\protect\cite{NegC-theo}
does not suffer of this limitation.
\end{thebibliography}
\end{document}